\documentclass{IEEEtran}
\usepackage{mathtools}
\usepackage{amsmath}
\usepackage{blindtext}
\usepackage{bbm}
\usepackage{comment}
\usepackage{graphicx}
\usepackage[ruled,vlined]{algorithm2e}
\newtheorem{lemma}{Lemma}
\usepackage{amsmath,xcolor,amssymb,multirow,cite,epsfig,graphicx,graphics}
\renewcommand{\baselinestretch}{1.2}

\begin{document}
\title{On Greedy Routing in Dynamic UAV Networks}
\vspace{30pt}

\author{Mehrdad Khaledi, 
Arnau Rovira-Sugranes, 
Fatemeh Afghah, 
and~Abolfazl Razi\\
School of Informatics, Computing and Cyber Systems, Northern Arizona University,\\
Email:\{mehrdad.khaledi,ar2832,fatemeh.afghah,abolfazl.razi\}@nau.edu}


\maketitle

\begin{abstract}

Unmanned aerial vehicles (UAVs), commonly known as drones, are becoming increasingly popular for various applications. 
Freely flying drones create highly dynamic environments, where conventional routing algorithms which rely on stationary network contact graphs fail to perform efficiently. Also, link establishment through exploring optimal paths using hello messages (as is used in AODV algorithm) deems extremely inefficient and costly for rapidly changing network topologies. 

In this paper, we present a distance-based greedy routing algorithm for UAV networks solely based on UAVs' local observations of their surrounding subnetwork. Thereby, neither a central decision maker nor a time consuming route setup and maintenance mechanism is required. 
To evaluate the proposed method, we derive an analytical bound for the expected number of hops that a packet traverses. Also, we find the expected end-to-end distance traveled by each packet as well as the probability of successful delivery. 
The simulation results verify the accuracy of the developed analytical expressions and show considerable improvement compared to centralized shortest path routing algorithms.
\end{abstract}
\vspace{-0.2 in}
\IEEEpeerreviewmaketitle

\section{Introduction}
Unmanned Aerial Vehicles (UAV) have recently attracted significant interest in many civilian, commercial and military applications. 
The popularity of UAVs emanate from their low cost, rapid deployment, and  ability to fly above obstacles. 
It is anticipated that the global market revenue of UAVs reach \$11.2 billion by 2020 \cite{Gartner_UAV}. 
Some applications require deployment of a large number of UAVs to complete designated tasks \cite{Razi_Asilomar17,Mousavi_INFOCOM18,Afghah_ACC18}. A network of autonomous UAVs form a flying Ad-hoc network (FANET), which provides more sensing and actuation capabilities. 
However, it also poses many challenges in designing networking protocols. 

One of the main challenges in FANETs is the high degree of mobility which causes frequent changes in network topology. As such, conventional communication protocols face a considerable performance degradation \cite{Razi_WiSEE,Razi_CCWC,Springer}. 
In routing, for instance, link breakage and frequent topology changes can lead to packet loss, excessive re-transmissions and eventually increased delay. 
 
For environments with low degrees of mobility or with geographically confined movements, several routing protocols have been proposed recently. For instance, in vehicular Ad-Hoc networks (VANETs), a routing protocol based on the Dijkstra's algorithm \cite{veh} and an optimized multicast routing protocol \cite{veh1} have been developed. In \cite{veh2}, a stable routing protocol is presented which finds the most stable path by considering velocity, direction and the link expiration time using fuzzy logic. In addition, learning-based routing protocols (e.g. Q-routing) have been proposed to learn the link states based on the current transmission experience \cite{choi1996predictiveQrouting}. Such algorithms rely on the assumption of low speed vehicles moving in a confined two dimensional space with limited and predefined movement patterns dictated by obstacles and roads. These assumptions, however, are not realistic in FANETs with UAVs moving freely in space with potentially high speeds.

For dynamic UAV networks, however, there have been few efforts on routing algorithm design. For instance, the authors in \cite{Gankhuyag17}, presented the RARP protocol which utilizes GPS information to estimate the duration of a path. This algorithm is based on AODV~\cite{perkins1999ad} and requires a route setup phase before transmission. To avoid repeated route setups the algorithm assumes that nodes keep their current movement patterns for some period of time which is a typical assumption in mobile Ad-Hoc networks (MANETs). This assumption is not valid for network of autonomous UAVs. This protocol is also evaluated using a mobility model primarily developed for low-speed terrestrial nodes, hence not suited for freely flying drones. Another recently proposed routing protocol for UAV networks is the Predictive-OLSR \cite{Rosati16} which utilizes GPS coordinates to estimate the quality of a link. However, they adopt a restrictive assumption that only the source node is mobile. 

In this paper, we present a distance greedy routing algorithm for dynamic UAV networks. This algorithm relies on local forwarding decisions and does not require a route setup phase. This low-complexity algorithm imposes no additional signaling overhead to the system, hence well suited to dynamic UAV networks. In order to analyze the performance of the greedy routing, we derive analytical lower and upper bounds for the expected number of hops that a packet traverses. Using this result, we find the expected end-to-end distance that a packets travels using the greedy algorithm, which can be a ground to optimize distance-based performance metrics such as end-to-end delay and transmission power consumption. 
We also derive the probability of successful delivery which is the probability that all intermediate nodes along the path can forward the packet to a node closer to 
the destination. Finally, simulation results verify the accuracy of the derived analytical results and also indicate superior performance of the greedy routing compared to the conventional shortest path routing based on Dijkstra's algorithm~\cite{Dijkstra59}.

\vspace{-0.1 in}
\section{System Model}\label{sysModel} 
We consider a FANET with $N$ UAV nodes ($n_1,n_2,\dots,n_N$), distributed uniformly in a $L\times L$ rectangular area. Let $R$ denote the radius of the circular communication range of a UAV node, then the set of neighbors for node $n_i$ is defined as:
\begin{equation*}
S_i= \Big\{n_j: d_{ij}=\sqrt[]{(x_i-x_j)^2+(y_i-y_j)^2}\leq R\Big\}
\end{equation*}
\noindent where $d_{ij}$ denotes the Euclidean distance between nodes $n_i$ and $n_j$.
Also, we use the popular mobility model for UAV nodes which integrates linear and circular motions \cite{mobility}. Transition between two mobility modes occur based on an underlying Markov process with adjustable transition probabilities. 
The mobility parameters include speed and direction of linear motion, and initial phase, angular speed, and radius of circular motion. Speed and radii  are drawn from exponential distributions and the rest of parameters are uniformly distributed. The parameters are initialized when transition occurs between the two states and remain constant until the next transition.

UAV nodes do not keep track of entire network topology, but each node needs to know the location of its neighbors. We assume the anticipated locations of neighbors can be predicted by a node, either through a model-based motion trajectory prediction method \cite{Aoude11} or exploiting online path-planning information by UAVs \cite{Goez16}. Also, since the algorithm works based on the remaining distance to the destination, the source node needs to know the location of the destination. This information can be embedded into the packet to inform the intermediate nodes about the destination's location.

\vspace{-0.1 in}
\section{The Distance Greedy Routing Algorithm}\label{routingAlg}   
Consider a source node $s$ who wants to send a packet to a destination node $t$ which is located at distance $D$. The distance greedy algorithm works based on a simple forwarding rule. At each step, $i$, the packet is passed by the current node $n_i$ to a neighbor node $n_j$ which is closest to the destination, i.e. $n_{i+1}=\underset{n_j \in S_i}{\text{argmin }} d_{j,t}$. 
The packet is passed only if the next node makes a progress, (i.e. if the next node is closer to the destination than the current node: $d_{i+1,t}<d_{i,t}$). If such a neighbor does not exist the current session fails and the packet journey is re-initiated. This algorithm continues until the packet is delivered to the destination. This constraint is required to ensure a loop-free path towards the destination. In this way, we guarantee a progress at each step provided that there is at least one neighbor in the \textit{progress area}. 

\begin{figure}[t]
	\centering
	\includegraphics[width=.85\columnwidth]{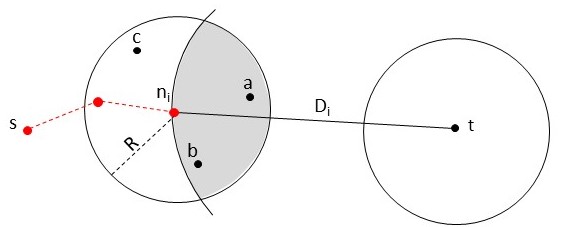}
	\caption{\footnotesize The intermediate node, $n_i$, forwards the packet to a neighbor which is closer to the destination, t, than its other neighbors ($n_{i+1}=a$ here). The shaded area called \textit{progress area} which is the valid locations for neighbors that have lower distances to the destination than the current distance, $D_i$.}
	\vspace{-0.2in}
    \label{fig:ProgressArea}
\end{figure}

Fig.~\ref{fig:ProgressArea}, represents one iteration of this algorithm. The shaded area represent the valid locations for the next node. 
This \textit{progress area} is the intersection of two circles centered at $n_i$ and $t$ with radii $R$ and $D_{i}$, respectively. Here, $D_i$ is the remaining distance to the destination, once the packet reaches node $n_i$. The algorithm chooses a node which makes the highest progress towards the destination (here $n_{i+i}=a$).

\section{Analysis of The Greedy Routing Algorithm}\label{analysis}

In this section, we evaluate the performance of the greedy routing algorithm 
through several steps.
%
%
%
%
We first bound the expected number of hops a packet travels from source to destination. Next, using the results we find the expected end-to-end distance traversed by the greedy algorithm. Finally, we find the probability of successful delivery in terms of transmission range, number of nodes and the size of the area through finding the probability of having at least one node in the \textit{progress area} throughout the selected path.

\subsection{Analysis of The Number of Hops}\label{analysisHops}

In order to bound the expected number of hops for sending a packet from source to destination, we first find the probability distribution of the progress made at each hop. 


Let $A_{D_i}(R,D_i)$ denote the area of the shaded region in Fig.~\ref{fig:ProgressArea} which is the intersection of two circles with centers at $n_i$, $t$ and with radii $R$ and $D_i$, respectively. Also, let $X_i$ be a random variable representing the remaining distance to the destination at the $i^{th}$ hop. In fact, $X_i$ represents the distance from destination $t$ to its closest node in the shaded area. The probability of $X_i$ being at least $x$ equals the probability that there are no nodes in the area $A_{D_i}(R,x)$ which is the overlap of two circles with centers at $n_i$ and $t$, which are at distance $D_i$ of each other, and with radii $R$ and $x$, respectively. Since, nodes are distributed uniformly, the number of nodes in any region with area $A$ follows a Binomial distribution with $N$ trials and success probability of $\frac{A}{L^2}$. Thus, we find
\begin{equation}
P[X_i\geq x]= \bigg(1-\frac{A_{D_i}(R,x)}{L^2}\bigg)^N,
\end{equation}
\noindent where $D_i-R\leq x < D_i$ and we can use geometric analysis to find the area $A_{D_i}(R,x)$ as
\begin{multline}\label{eq:overlapArea}
A_{D_i}(R,x)=R^2 \cos^{-1}\Big(\frac{D_i^2+R^2-x^2}{2D_iR} \Big) \\
 + x^2 \cos^{-1}\Big(\frac{D_i^2+x^2-R^2}{2D_ix} \Big) \\
 - \frac{1}{2}\sqrt[]{(R-D_i+x)(D_i-R+x)(D_i+R-x)(D_i+R+x)}.
\end{multline}
Now, we can find the probability distribution of the progress made at each hop. Let $Y_i=D_i-X_i$ denote the progress the $i^{th}$ hop, we have:
\begin{align} \label{eq:CDFProgress}
P[Y_i\leq y]= 
\begin{cases}
0 &y < 0 \\
\Big(1-\frac{A_{D_i}(R,D-y)}{L^2}\Big)^N&0 \leq y\leq R\\
1 &y> R
\end{cases}
\end{align}
The probability density function (PDF) of $Y_i$ can be computed by taking the derivative of its distribution function in (\ref{eq:CDFProgress}). Note that there is a discontinuity point at $Y_i=0$, therefore we can write the PDF as:
\begin{equation}\label{eq:PDFProgress}
f_{Y_i}(y)= P[Y_i\neq 0]f_{Y_i}^c(y|Y_i\neq 0) + P[Y_i=0]\delta (y),
\end{equation}
\noindent where $f_{Y_i}^c(y|Y_i\neq 0)$ denotes the continuous part conditioned on progress, which is the derivative of the distribution function for $Y_i$ between 0 and $R$. Knowing the PDF, we can find the expected progress at the $i^{th}$ hop as follows:
\begin{equation*}
   \mathbb{E}[Y_i] = \int_{0}^{R} y \;f_{Y_i}(y|Y_i\neq 0) \;dy.
\end{equation*}
\noindent Using integration by parts, we have:
\begin{eqnarray}\label{eq:ExpectedProgress}
   \nonumber \mathbb{E}[Y_i] &=& y \; \Big(1-\frac{A_{D_i}(R,D_i-y)}{L^2}\Big)^N \bigg]_{0}^{R} \\
 \nonumber  &-& \int_{0}^{R} \Big(1-\frac{A_{D_i}(R,D_i-y)}{L^2}\Big)^N) \;dy \\
  &=& R - \int_{0}^{R} \Big(1-\frac{A_{D_i}(R,D_i-y)}{L^2}\Big)^N) \;dy,
\end{eqnarray}
\noindent where the first term equals $R$ since $A_{D_i}(R,D_i-R)=0$ as there is no intersection between two circles with centers at $n_i$, $t$, which are at distance $D_i$ of each other, and with radii $R$ and $D_i-R$, respectively.  

Now that we have the average progress at each hop, we find the number of hops that a packet traverses to reach a destination located at distance $D$
of the source node. The number of hops is of this form; $n=m+1$ where $m$ is the number of hops needed to reach the communication range of the destination. That means, the first $m$ hops takes the packet to the destination's communication range where there is only one hop left to the destination node. We have:
\begin{equation}\label{eq:numberOfHops}
\sum_{i=1}^{m-1} Y_i < D-R \leq \sum_{i=1}^{m} Y_i.
\end{equation}

It should be noted that $m$ is a stopping time step with respect to the sequence $Y_i$. That means, at time $m$ we have enough information to stop and we do not need any future information to decide. For a special case of stopping times when the sequence of random variables are independent and identically distributed (i.i.d.), we can utilize the Wald's equation~\cite{Wald45} to find the sum of random variables up to time $m$.

\begin{lemma}[Wald's Equation~\cite{Wald45}]\label{lemma:Wald}
If $\tau$ is a stopping time with respect to an i.i.d. sequence $\{X_i : i\geq 1\}$, and if $\mathbb{E}[\tau]<\infty$ and $\mathbb{E}[X] < \infty$, then
\begin{equation*}
 \mathbb{E}\bigg[\sum_{i=1}^{\tau} X_i\bigg] = \mathbb{E}[\tau]\;\mathbb{E}[X].
\end{equation*}
\end{lemma}

In our case, however, the sequence $Y_i$ is not i.i.d., therefore, we cannot directly use the Wald's equation. For this reason, we first find the number of hops using some i.i.d. random variables $Z_i$. Next, we replace $Z_i$ with i.i.d. random variables that upper bound and lower bound $Y_i$. Thereby, we conclude about the bounds on the number of hops a packet travels. Using Lemma~\ref{lemma:Wald} for i.i.d. random variables $Z_i$, we have

\begin{equation}\label{eq:tempLB}
 \mathbb{E}\bigg[\sum_{i=1}^{m} Z_i\bigg] = \mathbb{E}[m]\;\mathbb{E}[Z].
\end{equation}
\noindent from the inequality in (\ref{eq:numberOfHops}), we have:
\begin{equation}\label{eq:tempLB2}
 \mathbb{E}\bigg[\sum_{i=1}^{m} Z_i\bigg] \geq D-R. 
\end{equation}
\noindent Combining (\ref{eq:tempLB}) and (\ref{eq:tempLB2}) we get:
\begin{equation}\label{eq:LBforZ}
 \mathbb{E}[m] \geq \frac{D-R}{\mathbb{E}[Z]} .
\end{equation}

To find an upper bound for $\mathbb{E}[m]$ we use the left inequality in (\ref{eq:numberOfHops}) and the fact that the progress at each hop is at most $R$,
\begin{equation*}
 \sum_{i=1}^{m} Z_i \leq \sum_{i=1}^{m-1} Z_i + R < D.
\end{equation*}
\noindent taking expectation we have 
\begin{equation*}
 \mathbb{E}\bigg[\sum_{i=1}^{m} Z_i\bigg] < D .
\end{equation*}
\noindent using (\ref{eq:tempLB}), we get:
\begin{equation}\label{eq:UBforZ}
 \mathbb{E}[m] < \frac{D}{\mathbb{E}[Z]} 
\end{equation}

Now, utilizing (\ref{eq:UBforZ}) and (\ref{eq:LBforZ}) the expected number of hops $\mathbb{E}[n]=\mathbb{E}[m]+1$ can be bounded as follows
\begin{equation}\label{eq:BoundforZ}
 \frac{D-R}{\mathbb{E}[Z]} +1\leq \mathbb{E}[n] < \frac{D}{\mathbb{E}[Z]} +1.
\end{equation}

As mentioned earlier the random variables $Y_i$ are not i.i.d. and we need to bound them using i.i.d. random variables to be able to use the result in (\ref{eq:BoundforZ}). 

For this purpose, let us define the progress at each hop as a function of the remaining distance to the destination, as $Y(D)$. It is worth noting that the $Y(D)$ is a non-decreasing function of the remaining distance $D$. More precisely, for $D\geq D'$ we have $P[Y(D)>y] \geq P[Y(D')>y]$. Intuitively, if there is more distance to the destination it is more likely that we make more progress than the case of having less distance to the destination. This can be shown formally using (\ref{eq:CDFProgress}), as follows

\begin{eqnarray}\label{eq:Monontonicity}
 P[Y(D)>y] = 1- \Big(1-\frac{A_D(R,D-y)}{L^2}\Big)^N \geq \\
 \nonumber 1- \Big(1-\frac{A_{D'}(R,D'-y)}{L^2}\Big)^N = P[Y(D')>y],
\end{eqnarray}
\noindent where the inequality follows from the fact that if $D\geq D'$ we have $A_{D}(R,D-y)\geq A_{D'}(R,D'-y)$. 

Now, observe that at any of the $m$ hops, the distance between the intermediate node and the destination is between $R$ and $D$. Using (\ref{eq:Monontonicity}), we get
\begin{eqnarray}
P[Y(R)>y] \leq P[Y_i>y]\leq P[Y(D)>y]. \\ 
\nonumber \text{for}\; i=1,\cdots,m 
\end{eqnarray}

Thus, in (\ref{eq:BoundforZ}) we can replace $Z_i$ with i.i.d. random variables $Y(D)$ to find a lower bound for the number of hops. Similarly, we can use i.i.d. random variables $Y(R)$ to find an upper bound.
\begin{equation}\label{eq:BoundHops}
 (1_{\{D\geq R\}})\;\cdot \frac{D-R}{\mathbb{E}[Y(D)]} +1\leq \mathbb{E}[n] < \frac{D}{\mathbb{E}[Y(R)]} +1,
\end{equation}
\noindent where $\mathbb{E}[Y(D)]$ can be computed using (\ref{eq:ExpectedProgress}) and $1_{\{D\geq R\}}$ is an indicator to account for the case where source and destination are immediate neighbors.

\subsection{Analysis of The End-to-End Delay}\label{analysisDelay}
In this section, we analyze the total distance a packet travels from source to destination using the distance greedy routing. Considering the delay is proportional to the distance, this will give us the end-to-end delay metric.

\begin{figure}[t]
	\centering
	\includegraphics[width=.85\columnwidth]{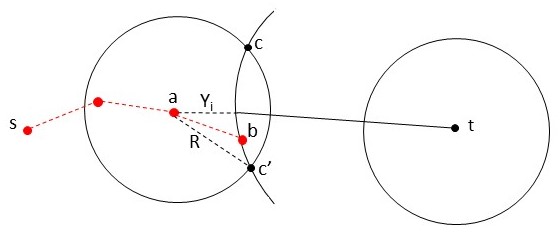}
	\caption{\footnotesize The length of line $\bar{ab}$ can range between $Y_i$, the progress made at the $i^{th}$ hop, and the transmission range R.}
	\vspace{-0.05in}
    \label{fig:DistanceTraveled}
\end{figure}

Consider the forwarding scenario at the $i^{th}$ hop, depicted in Fig.~\ref{fig:DistanceTraveled}, where node $b$ has been chosen by the intermediate node $a$ and we make a progress of $Y_i$ towards the destination. We want to find the distance traveled at the $i^{th}$ hop (i.e. the length of the line $\bar{ab}$ in Fig.~\ref{fig:DistanceTraveled}), which we denote it by $W_i$. Given the progress $Y_i$, we know that node $b$ should be on the arc $cc'$. Considering the fact that nodes are uniformly distributed, node $b$ can be anywhere on the arc $cc'$ with equal probability. Also, the distance of any node on the arc $cc'$ from the transmitting node $a$ ranges between $Y_i$ and $R$. Therefore, we can roughly estimate that the distance from node $a$ to node $b$ is uniformly distributed between $[Y_i, R]$. That is $W_i\sim U[Y_i, R]$ for the $i^{th}$ hop. 

Now, we can find the expected end-to-end distance traveled by a packet as
\begin{equation}\label{eq:TotalDistance}
 \mathbb{E}\bigg[\sum_{i=1}^{n} W_i\bigg] =\sum_{i=1}^{n} \mathbb{E} [W_i]= \sum_{i=1}^{n} \frac{Y_i + R}{2}=\frac{1}{2} (D + \mathbb{E}[n]R),
\end{equation}

\noindent where $n$ is the number of hops whose expectation is characterized in (\ref{eq:BoundHops}). Also, we have used the fact that the progresses at each hop sum up to the distance between source and destination, $\sum_{i=1}^{n} Y_i = D$.

\subsection{Analysis of Network Density for Successful Delivery}\label{AnalysisPSuccess}

In the greedy routing algorithm, we assume that there is always a neighbor to forward the packet towards the destination and thereby we ignore the possibility of the packet reaching an isolated node (more precisely, a node without any progress-making neighbors). In fact, if the network is dense enough or if the transmission range of nodes are large enough, such a situation can be avoided. 

In this section, we find the probability of successful delivery\footnote{It is worth noting that by successful delivery we mean the packet travels from source to destination without facing an isolated node.}. First, we find the probability of a node being isolated which equals the probability of having no node in its \textit{progress area}. To simplify the analysis, we estimate the \textit{progress area} of a node by the half of its transmission region which faces the destination. Then, we can write the probability of a node's isolation, $P_{iso}$, as
\begin{equation}\label{eq:isolation}
 P_{iso} \approx \Big(1-\frac{\frac{\pi R^2}{2} }{L^2}\Big)^{N-1} 
\end{equation}

\noindent we can now solve (\ref{eq:isolation}) for $R$ and find the minimum transmission range such that the probability of node isolation is less than $\epsilon$
\begin{equation}\label{eq:Risolation}
 R> \sqrt[]{\frac{2}{\pi}} \; \;\sqrt[]{L^2 +L^2( \epsilon^{\frac{1}{N-1}})}
\end{equation}

The probability of success equals the probability of no node along the path being isolated which is $(1-P_{iso})^n$, where $n$ denotes the number of hopes that is bounded in expectation by (\ref{eq:BoundHops})
\begin{equation}\label{eq:success}
 P_{success} \approx \Bigg( 1- \Big(1-\frac{\pi R^2 }{2L^2}\Big)^{N-1} \Bigg)^n
\end{equation}

\section{Simulation Results}\label{simResults}

To test simulations results and prove the efficacy of the analysis work, random networks are generated using uniform distributions for the initial real positions in a $L\times L$ grid. We use the mobility model explained in Section II 
to generate motion trajectories for $N$ nodes. We use the actual positions for all nodes when quantifying the performance metric, but use the predicted positions when finding the optimal path. The predicted locations are the actual locations mixed with Normally distributed prediction noise of variance $\sigma_N^2=10$.

We use dynamic contact graph by making connections between nodes with pairwise distances below $R$. 
The rest of simulation parameters include number of nodes: $N=10$, the grid size: $L=10~km$, communication range: $R=5~km$, average node velocities: $\bar{v}=50 m/sec$, and average waiting time: $\bar{w}=20$, unless specified otherwise. Also the transition probability between the circular and linear motions is 20\%. Finally, we note that for all figures, we take average over $100$ runs of the algorithm with different initializations.

We first, verify the accuracy of the derived upper and lower bounds for two important performance metrics, namely the end-to-end delay per packet and the probability of success.  

\begin{figure}[t]
	\centering
	\includegraphics[width=.9\columnwidth]{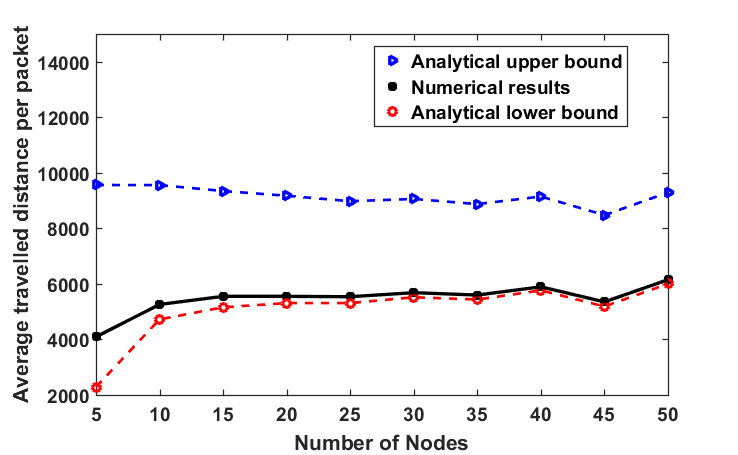}
	\caption{\footnotesize Total travel distance from source to destination per packet vs the number of nodes in the network ($N$): simulation results are compared against analytical lower and upper bounds.}
	\vspace{-0.15 in}
    \label{fig_D_N}
\end{figure}

In Fig.\ref{fig_D_N}, we present the expected end-to-end distance per packet vs $N$. The upper and lower bounds are presented based on \ref{eq:TotalDistance}, where the bounds on the expected number of hops, $\mathbb{E}[n]$ is obtained from (\ref{eq:BoundHops}). 
We note that the lower bound is tighter, which is due to the tightness of the lower bound in (\ref{eq:BoundHops}). 
The fluctuation in the results is due to the average distance between the source and destination (D), which is a probabilistic value.

\begin{figure}[t]
	\centering
	\includegraphics[width=.9\columnwidth]{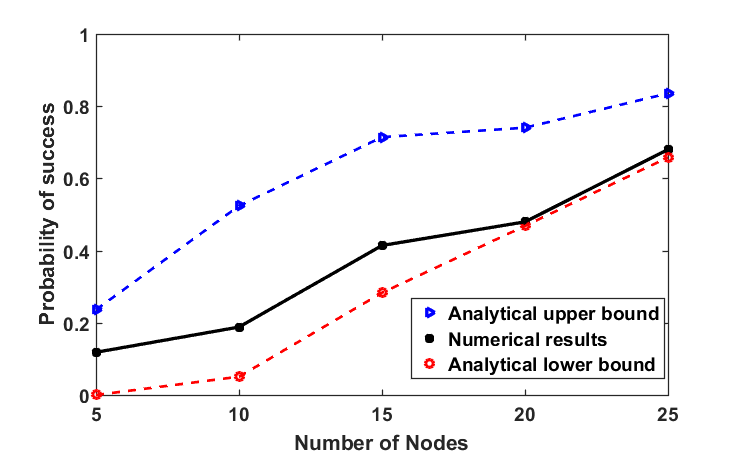}
	\caption{\footnotesize Probability of delivery success versus the number of nodes: comparison between the simulation results and analytical lower and upper bounds.}
	\vspace{-0.15in}
    \label{fig_psuccess_N}
\end{figure}

Another important performance indicator of the proposed algorithm is the probability of success, which means possibility of progress at all intermediate nodes (having at least one node in the current node's \textit{progress area}), as characterized in (\ref{eq:success}) based on the average number of hops per packet in (\ref{eq:BoundHops}).
Fig \ref{fig_psuccess_N} suggests that as we increase the number of nodes, the network density increases and therefore the probability of getting stuck in an intermediate node with empty \textit{progress area} diminishes. Similar to Fig. \ref{fig_D_N}, the obtained lower bound is more accurate. 

\begin{figure}[t]
	\centering
	\includegraphics[width=.9\columnwidth]{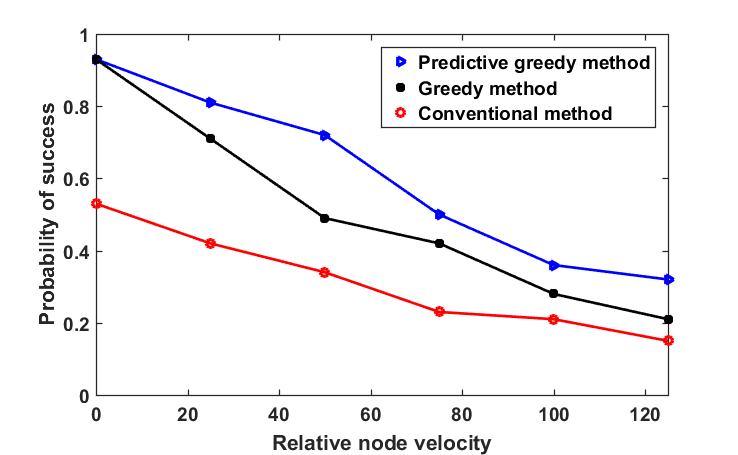}
	\caption{\footnotesize Probability of delivery success for predictive greedy algorithm, static greedy algorithm and conventional Dijkstra's algorithm.}
	\vspace{-0.15in}
    \label{figcomp}
\end{figure}

Now, we compare the performance of the proposed greedy algorithm with the conventional Dijkstra's shortest path algorithm in Fig. \ref{figcomp}. We also evaluate the proposed algorithm with and without including predictive location information under different average node velocities. The results show that the probability of delivery success for the greedy method is higher than that of the conventional shortest path algorithm consistently for all average node velocities. 
Also, the predictive greedy method outperforms the static greedy algorithm, which shows including predictive location information decreases the probability of selecting nodes with empty \textit{progress area}, as was expected. 
Finally, when the network is more dynamic, more nodes are subject to getting out of the communication ranges of their neighbors, and hence the probability of success declines.

\begin{figure}[t]
	\centering
	\includegraphics[width=.85\columnwidth]{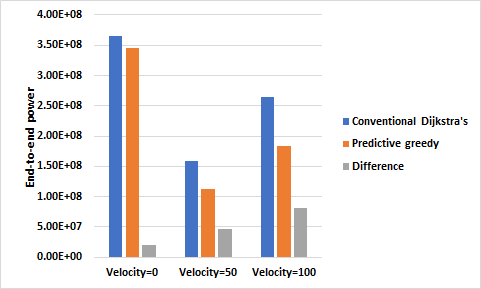}
	\caption{\footnotesize End-to-end power comparison between the conventional Dijkstra's algorithm and the predictive greedy algorithm.}
	\vspace{-0.15in}
    \label{barplot}
\end{figure}

Lastly, in Fig. \ref{barplot} we present the average power consumption per packet to complete the path for the proposed predictive greedy algorithm and the standard Dijkstra's algorithm without including predictive information in order to show the practical utility of the proposed method. Since the power consumption is proportional to the sum of all link distances squared, considering predictive information provides a significant gain for our suboptimal algorithm. This gain is higher for more dynamic networks, since the inclusion of predictive locations is more beneficial for higher average node velocities.

\section{Conclusion}\label{conclusions}
In this paper, we studied the routing problem in dynamic UAV networks. Prior approaches fail to perform well in such dynamic environments due to the requirement of maintaining information about the network topology or using frequent route-establishment phases. We studied an agile distance-greedy routing algorithm which is low-complexity and the intermediate nodes take decisions solely based on the predicted locations of their neighbors. This algorithm is fully distributed and incorporates predicted locations into the algorithm, hence outperforms the centralized shortest algorithm.  
We characterized the number of hops, the probability of success delivery and the expected distance a packet travels based on system parameters. We plan to characterize the impact of the prediction uncertainty on the optimality of the selected path as an extension of this work.
\vspace{-0.15in}
\bibliography{main}
\bibliographystyle{IEEEtran}

\end{document}